\documentclass[aps,prd,twocolumn,amsmath,nofootinbib,superscriptaddress]{revtex4-2}
\usepackage{graphicx}
\usepackage{float}
\usepackage{epstopdf}
\usepackage{epsfig}
\usepackage{amsmath}
\usepackage{color}
\usepackage{amssymb}   % \gtrsim, \geqslant, etc etc: see amsguide.ps
\usepackage{amsfonts}  % \mathfrak and \mathbb{x} (Blackboard bold)
\usepackage{amsbsy}    % \pmb and \boldsymbol
\usepackage{pstricks}
\usepackage{slashed}
\usepackage{dcolumn}
\usepackage{xparse}
\usepackage{bm}
\usepackage{hyperref}
\usepackage{soul}
\usepackage{bbm}
\usepackage[T1]{fontenc}
\newcommand{\al}{\alpha}
\newcommand{\be}{\beta}

\newcommand{\beq}{\begin{equation}}
\newcommand{\eeq}{\end{equation}}
\newcommand{\ba}{\begin{array}}
\newcommand{\ea}{\end{array}}
\newcommand{\bea}{\begin{align}}
\newcommand{\eea}{\end{align}}
\newcommand{\bi}{\begin{itemize}}
\newcommand{\ei}{\end{itemize}}
\newcommand{\ben}{\begin{enumerate}}
\newcommand{\een}{\end{enumerate}}
\newcommand{\bc}{\begin{center}}
\newcommand{\ec}{\end{center}}
\newcommand{\bl}{\begin{flushleft}}
\newcommand{\el}{\end{flushleft}}
\newcommand{\br}{\begin{flushright}}
\newcommand{\er}{\end{flushright}}

\newcommand{\nn}{\nonumber \\}
\newcommand\Eqn[1]{Eq.~(\ref{#1})}  % includes ``Eq.'' in front
\newcommand{\ie}{{i.e.}}
\newcommand{\eg}{{e.g.}}
\newcommand{\GeV}{{\rm GeV}}

\newcommand\comment[1]{ \hbox{[{\it Comment suppressed here.}\/]} }
\newcommand\hide[1]{}
\newcommand{\skipover}[1]{}

\begin{document}
%\captionsetup{justification=raggedright}
%\title{Pion gravitational form factor in the contact model}
%\title{A glimpse into pion gravitational form factor with Dyson-Schwinger equation %approach}
\title{QCD anomalies in electromagnetic processes: A solution to the \texorpdfstring{$\gamma\to3\pi$}{} puzzle}

\author{Zanbin Xing}%
\email{xingzb@mail.nankai.edu.cn}
\affiliation{School of Physics, Nankai University, Tianjin 300071, China}
\author{Hao Dang}%
\email{2011707@mail.nankai.edu.cn}
\affiliation{School of Physics, Nankai University, Tianjin 300071, China}
\author{M. Atif Sultan}%
\email{atifsultan.chep@pu.edu.pk}
\affiliation{School of Physics, Nankai University, Tianjin 300071, China}
\affiliation{  Centre  For  High  Energy  Physics,  University  of  the  Punjab,  Lahore  (54590),  Pakistan}
\author{Kh\'epani Raya}
\email{khepani.raya@dci.uhu.es}
\affiliation{Department of Integrated Sciences and Center for Advanced Studies in Physics, Mathematics and Computation, University of Huelva, E-21071 Huelva, Spain.}
\author{Lei Chang}%
\email{leichang@nankai.edu.cn}
\affiliation{School of Physics, Nankai University, Tianjin 300071, China}

\date{\today}
\begin{abstract}
%It seems that there is a discrepancy between theory and experiments for the $\gamma\to3\pi$ anomaly. 
In this work, the $\gamma\to3\pi$ form factor is calculated within the Dyson-Schwinger equations framework using a contact interaction model within the so-called modified rainbow ladder truncation. The present calculation takes into account the pseudovector component in the pion Bethe-Salpeter amplitude (BSA) and $\pi-\pi$ scattering effects, producing a $\gamma\to3\pi$ anomaly which is $1+6\mathcal{R}_\pi^2$ larger than the low energy prediction. Here $\mathcal{R_\pi}$ is the relative ratio of the pseudovector and pseudoscalar components in the pion BSA; with our parameters input, this correction raises the $\gamma\to3\pi$ anomaly by around $10\%$. The main outcome of this work is the unveiling of the origin of such correction, which could be a possible explanation of the discrepancy between the existing experimental data and the low energy prediction.  Moreover, it is  highlighted how the magnitude of the anomaly is affected in effective theories that require an irremovable ultraviolet cutoff. We find that for both the anomalous processes $\pi\to2\gamma$ and $\gamma\to 3\pi$, the missing contribution to the anomaly can be compensated by the additional structures related with the quark anomalous magnetic moment. %generated by the modified rainbow ladder truncation uniformly.
\end{abstract}

%In this work the $\gamma\to3\pi$ form factor is calculated within the Dyson-Schwinger framework using the contact interaction model with the so-called modified rainbow ladder truncation. After a careful calculation in a complete and self consistent frame work, a correction to the low energy theorem is found. As a combined result of the pseudo vector component in the pion Bethe-Salpeter amplitudes and the $\pi-\pi$ scattering effect, the calculated $\gamma\to3\pi$ anomaly is $1+6\mathcal{R}_\pi^2$ larger than the low energy prediction, where $\mathcal{R_\pi}$ is the relative ratio of the pseudo vector and pseudo scalar components in the pion Bethe-Salpeter amplitudes. With our parameter input, this correction raises the $\gamma\to3\pi$ anomaly by around $10\%$. This correction is the primary result of this work and could be a possible explanation of the discrepancy between existing experimental data and low energy prediction. In addition, since the cutoff in contact model can not be removed, the anomaly is incomplete because of the cutoff. We find that for both the anomalous processes $\pi\to2\gamma$ and $\gamma\to 3\pi$, the missing contribution to the anomaly can be compensated by the additional structures generated by the modified rainbow ladder truncation uniformly.

\maketitle
\section{introduction}
The anomaly structure of Quantum Chromodynamics (QCD) can be investigated through a series of anomalous electromagnetic processes which involve an odd number of pseudoscalar mesons. The most famous one, perhaps, is the decay of neutral pion into two photons, which has close relation to the chiral anomaly discovered by Adler, Bell and Jackiw in 1969 \cite{Bell:1969ts,Adler:1969gk}. A basic result of QCD's quantization is that such processes occur in the chiral limit, $m_{\pi}=0$. In this work we focus on the process $\gamma\to3\pi$, accessible to experiments, which includes three hadronic bound states and hence provides additional insights into QCD. Despite the complexity of strong interactions, the amplitudes for the $\pi\to2\gamma$ and $\gamma\to3\pi$ are elegantly connected according to the low energy theorem \cite{Adler:1971nq,Aviv:1971hq,Terentev:1971cso} in the low energy and chiral limit
\begin{equation}
    A^{\pi}_0=e f_\pi^{2} A^{3\pi}_0\,;
\end{equation}
here $A^{\pi}_0$ and $A^{3\pi}_0$ are the coupling constants associated with the $\pi\to2\gamma$ and $\gamma\to3\pi$ processes, respectively, and the value of these limit-case amplitudes can be accessed from the Wess–Zumino–Witten (WZW) action \cite{Wess:1971yu,Witten:1983tw}, which gives:
\begin{equation}
    A^{\pi}_0=\frac{N_c e^2}{12 \pi^2 f_\pi},\ A^{3\pi}_0=\frac{N_c e}{12 \pi^2 f_\pi^3}\,.
\end{equation}
Naturally, $N_c$ is the number of colors, $e$ is the elementary charge, and $f_\pi$ is the pion leptonic decay constant. 
For the $\pi\to2\gamma$ process, the extraction of $A^{\pi}$ from experimental result\,\cite{ParticleDataGroup:2022pth} is in good agreement with the low energy prediction when $N_c=3$, which is a strong evidence that quarks possess a new degree of freedom, color. On the other hand, there is a discrepancy between experiments and the low energy prediction for $A^{3\pi}$. The early experimental result\,\cite{Antipov:1986tp} is almost 40 years old,
\begin{equation}
    A^{3\pi}_{exp}=12.9\pm0.9\pm0.5\, \GeV^{-3},
\end{equation}
which is about $4/3$ larger compared to the low energy prediction 
\begin{equation}
    A^{3\pi}_{theo}=A^{3\pi}_0(f_\pi=0.092\GeV)=9.8\,\GeV^{-3}.
\end{equation}
Theoretical efforts have been made to explain this discrepancy, such as the incorporation of one loop and two loop corrections \cite{Hannah:2001ee,Bijnens:1989ff}; other works can be found in Refs.~\cite{Ametller:2001yk,Holstein:1995qj,Benic:2011rk,RuizArriola:1993sp,Ivanov:1995sf,Cotanch:2003xv,Niehus:2021iin}. However, the available data is not enough to verify the various theoretical results. The COMPASS experiment at CERN is currently conducting a precision experiment on $A^{3\pi}$, where the pion-photon scattering is mediated via the Primakoff effect using heavy nuclei as a target \cite{Ecker:2023qae}. 
%The prelimilary estimate by using the dispersive model \cite{Hoferichter:2012pm} is availaible
%\begin{equation}
    %A^{3\pi}_{exp}=10.3\pm0.1\pm0.6 \GeV^{-3}
%\end{equation}
Hopefully more accurate experimental results can be obtained in the near future and give more insights into the anomalies of QCD.\par
Motivated by the ongoing experiment and the historical missmatch with theory, we calculate $\gamma\to3\pi$ form factor in the formalism of Dyson-Schwinger equations (DSEs). In previous DSE exploration \cite{Alkofer:1995jx}, it is shown that the low energy theorem is already saturated by considering the leading structure, \ie, $i \gamma_5$, of the pion Bethe-Salpeter amplitude (BSA) and the generalized impulse approximation (GIA). However, there are limitations in two aspects. Firstly, the pion is both a quark-antiquark bound state and a (pseudo) Nambu-Goldstone boson of dynamical chiral symmetry breaking. Therefore, one should expect a  much richer structure for the pion BSA than merely the $i\gamma _{5}$ component\,\cite{Maris:1998hc}. And although these structures are, in principle, subdominant, its presence is required to satisfy crucial symmetries\,\cite{Maris:1997hd}, and its effects in different quantities can manifest themselves in a noticeable way. For example, the ultraviolet behavior of the pion electromagnetic form factor is dominated by the pseudovector component \cite{Maris:1998hc}. In this way, one cannot conclude that the contribution from non-leading structures of the BSAs will cancel out in the anomaly calculation either. Secondly, the amplitudes of $\gamma\to3\pi$ is not solely determined by the GIA, as Ref.~\cite{Bando:1993qy} has formally proved in the case of ladder approximation. Additional contributions from gluon exchanges in different channels of Mandelstam variables should be considered, which we call $\pi-\pi$ scattering contributions. The self consistent calculation of $\gamma\to3\pi$ form factor is presented in Ref.~\cite{Cotanch:2003xv}. The results indicate that the net contribution from the pion non-leading structure is not zero in the GIA diagram. The $\pi-\pi$ scattering diagram seems to cancel with the non-leading structure contribution in the GIA and leads to a result that is close to the low energy prediction.\par
Unfortunately, due to the complicated interaction employed therein, Ref.~\cite{Cotanch:2003xv} has presented only numerical results. In order to investigate in more detail the magnitude and origin of the anomaly, tracing down the numerical outcomes to the fundamental ingredients, herein we use the contact interaction (CI) model \cite{Frederico:1992np,Gutierrez-Guerrero:2010waf}, which exposes essential features of non-perturbative QCD, \ie, confinement and dynamical chiral symmetry breaking (DCSB). By utilizing the CI model, one is able to obtain rather simple expressions that can be employed to perform comparisons with the low energy prediction and other QCD-based models, in a practical manner. As will be detailed later, the CI model is non renormalizable, entailing that an ultraviolet cutoff playing a dynamical role must be introduced and cannot be removed; a fact that will involve additional subtleties in the calculation. On one hand, it is known that in effective theories the chiral anomaly related to the $\pi\to2\gamma$ transition is not completely reproduced, see, \eg, Refs.~\cite{Blin:1987hw,Alkofer:1992nh}. The missing part is a result of cutoff-dependent higher order contributions \cite{Alkofer:1992nh}. On the other hand, as demonstrated in Ref.~\cite{Dang:2023ysl}, under the so-called modified rainbow ladder (MRL) truncation \cite{Xing:2021dwe}, a quark anomalous magnetic moment (AMM) term in the quark photon vertex (QPV) emerges naturally. This term meets the mathematical requirements to be interpreted as beyond the cutoff corrections and\,\cite{Lepage:1997cs}, at the same time, has an intimate connection with DCSB, so its use in the $\pi\to2\gamma$ case would also be physically justified. With these ideas in mind, we therefore adopt the MRL truncation in the numerical calculation of $\gamma\to3\pi$ process.\par
This manuscript is organized as follows: in Sec.~\ref{sec::formalism}, we first introduce the notation and conventions necessary for the description of $\gamma\to3\pi$ process in the CI model under MRL truncation. %and describe how the AMM term of QPV plays an important role. 
In Sec.~\ref{sec::chiraltrace}, we compute the amplitude of the anomalous process, $\gamma\to3\pi$, by carefully performing the chiral trace and regularization, in order to avoid mathematical inconsistencies. In Sec.~\ref{sec::lowenergy}, we achieve the primary goal of this work by explicitly showing how the rich structures of the pion and the $\pi-\pi$ scattering effects contribute to the $\gamma\to3\pi$ anomaly. The numerical results produced under the MRL truncation are discussed in Sec.~\ref{sec::momentum}, before we close with a summary in Sec. ~\ref{sec::summary}.
%This manuscript is organized as follows: the description of $\gamma\to3\pi$ process in the formalism of DSEs is introduced in Sec.~\ref{sec::formalism}; since the amplitude of the $\gamma\to3\pi$ process is anomalous, careful computation should be performed to avoid mathematical inconsistency, this is discussed in Sec.~\ref{sec::chiraltrace}; the primary goal of this work is to show how the rich structures of pion and the $\pi-\pi$ scattering effect contribute to the $\gamma\to3\pi$ anomaly and is achieved in Sec.~\ref{sec::lowenergy}, the numerical results computed under MRL truncation is presented in Sec.~\ref{sec::momentum}; we close this manuscript with a summary.
\section{formalism}\label{sec::formalism}
\subsection{Amplitude for \texorpdfstring{$\gamma\to3\pi$}{} process}\label{sec::amplitude}
\begin{figure}[ht]
\includegraphics[width=8.6cm]{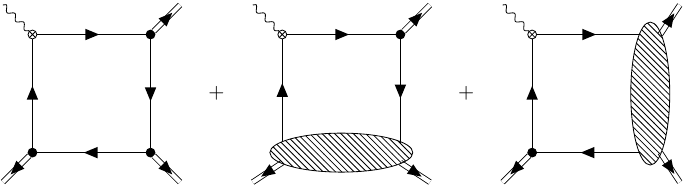}
\caption{A configuration of the diagrams contributing to the $\gamma\to3\pi$ process. The leftmost one is the box diagram, while the rest two diagrams are the corresponding $\pi-\pi$ scattering diagrams. The solid, double solid and wavy lines denote quark, the pion and the photon; the solid circle, crossed circle and shaded ellipse denote the pion BSA, the QPV and the self energy part of the $\pi-\pi$ scattering amplitude.}
\label{fig::amplitudes}
\end{figure}

Let us start by considering the anomalous process
\begin{equation}
    \gamma(Q)\to\pi^{+}(-P_2)\pi^{-}(-P_3)\pi^{0}(-P_4),
\end{equation} 
whose amplitude can be written as\footnote{We employ an Euclidean metric with $\{\gamma_\mu,\gamma_\nu\} = 2\delta_{\mu\nu}$; $\gamma_\mu^\dagger = \gamma_\mu$; $\gamma_5= \gamma_4\gamma_1\gamma_2\gamma_3$ so that $\text{tr}(\gamma_5\gamma_a\gamma_b\gamma_c\gamma_d)=-4\epsilon_{abcd}$; and $a \cdot b = \sum_{i}^{4} a_i b_i$. The isospin symmetry is considered herein.}:
\begin{equation}\label{eqn::totalamplitude}
    T^{3\pi}_\mu(s,t,u)=-\epsilon_{\mu P_2 P_3 P_4} A^{3\pi}(s,t,u),
\end{equation}
where the Mandelstam variables are $s=-(Q+P_2)^2$, $t=-(Q+P_3)^2$, $u=-(Q+P_4)^2$. All three pions are on-shell. The photon momentum is $Q=-(P_2+P_3+P_4)$ and $Q^2$ is related to the Mandelstam variables via 
\begin{equation}
    s+t+u=3 m^2_{\pi}-Q^2.
\end{equation}
The total amplitude $T^{3\pi}_\mu(s,t,u)$, according to the permutation of $\{P_2,+,s\}\leftrightarrow\{P_3,-,t\}\leftrightarrow\{P_4,0,u\}$, contains six configurations. In the formalism of DSEs, a complete and self-consistent calculation requires that each configuration be described by three diagrams. The first of them is box diagram, which is also called GIA in $\gamma\to3\pi$ process. This is shown in the left panel of Fig.\,\ref{fig::amplitudes}. The GIA is a good approximation in the calculation of the 3-body processes such as $\pi\to 2\gamma$ decay, producing in such case a triangle diagram. However, the box diagram resulting from the GIA is not sufficient to describe the 4-body process; one should also consider the $\pi-\pi$ scattering effects\,\cite{Bando:1993qy,Cotanch:2003xv}. The corresponding contributions are shown in the middle and right panel of Fig.\,\ref{fig::amplitudes}.

For simplicity, let us first consider one of the six configurations that contribute in this process (the rest is obtained from the aforementioned permutations). The graphical representation of this configuration is depicted in Fig.~\ref{fig::amplitudes}.
The expression for the box diagram is
\begin{eqnarray}\label{eqn::box}
T^{box}_\mu(s,t,u)\!\!\!&=&\!\!\!\text{tr}\!\int_{q}\Gamma_\mu(Q)S(q+P_2+P_4)\Gamma_\pi(P_2)S(q+P_4)\nn
\!\!\!&&\!\!\!\times\Gamma_\pi(P_4)S(q)\Gamma_\pi(P_3)S(q-P_3)\,,
\end{eqnarray}
and the corresponding $\pi-\pi$ scattering diagrams are given by
\begin{eqnarray}
\label{eqn::scat1}T^{scat1}_\mu(s,t,u)&=&\text{tr}\int_{q}\Gamma_\mu(Q)S(q+P_2+P_4)\Gamma_\pi(P_2)\nn
&&\times S(q+P_4)\Sigma^F(P_4,P_3)S(q-P_3)\,,\\
\label{eqn::scat2}T^{scat2}_\mu(s,t,u)&=&\text{tr}\int_{q}\Gamma_\mu(Q)S(q+P_2+P_4)\Sigma^F(P_2,P_4)\nn
&&\times S(q)\Gamma_\pi(P_3)S(q-P_3)\,,
\end{eqnarray}
where $\int_q\doteq\int\frac{d^4q}{(2\pi)^4}$ denotes a Poincar\'e invariant integration and the trace is over dirac space. 
$S(q)$ is the quark propagator with momentum $q$, $\Gamma_\pi(P)$ is the pion BSA with incoming pion momentum $P$, $\Gamma_\mu(Q)$ is the quark-photon vertex with incoming photon momentum $Q$ and $\Sigma^F(P_4,P_3)$ is the self-energy part of the $\pi-\pi$ scattering amplitude; $P_{3,4}$ are the momena of the two incoming pions. All these elements are further explained in Sec.~\ref{sec::ci}

Following \Eqn{eqn::totalamplitude}, The above Eqs.~(\ref{eqn::box}-\ref{eqn::scat2}) can be written in terms of scalar functions as %In analogy with total amplitude \Eqn{eqn::totalamplitude}, they can be parameterized as
\begin{eqnarray}
    T^{box}_\mu(s,t,u)&=&-\epsilon_{\mu P_2 P_3 P_4} f^{box}(s,t,u)\,,\\
    T^{scat1}_\mu(s,t,u)&=&-\epsilon_{\mu P_2 P_3 P_4} f^{scat1}(s,t,u)\,,\\
    T^{scat2}_\mu(s,t,u)&=&-\epsilon_{\mu P_2 P_3 P_4} f^{scat2}(s,t,u)\,.
\end{eqnarray}
According to charge conjugation symmetry, the following relations between the scalar functions hold:
\begin{eqnarray}
    f^{box}(s,t,u)&=&f^{box}(t,s,u)\,,\\
    f^{scat1}(s,t,u)&=&f^{scat2}(t,s,u)\,.
\end{eqnarray}
We then proceed by defining
\begin{equation}
    f^{scat}(s,t,u)\doteq f^{scat1}(s,t,u)+f^{scat2}(s,t,u)\,.
\end{equation}
Finally, the total form factor can be written as
\begin{equation}
    A^{3\pi}(s,t,u)= A^{box}(s,t,u)+A^{scat}(s,t,u)
\end{equation}
where $A^{\#}(s,t,u)$ (with $\#=\{box,scat\}$) is given by
 \begin{eqnarray}\label{eqn::ffs}
     A^{\#}(s,t,u)&=&[f^{\#}(s,t,u)+f^{\#}(s,u,t)+f^{\#}(u,t,s)]\nn
     & &\times (N_c) \times \left(\frac{Q_u+Q_d}{\sqrt{2}}\right)\times(\sqrt{2})^3\,.    
 \end{eqnarray}
The overall factor, in the second line of \Eqn{eqn::ffs}, contains three parts: $N_c=3$ comes from the trace over color space; $(\frac{Q_u+Q_d}{\sqrt{2}})$ is the flavor space factor, with $Q_{u,d}$ being the charges of $u$ and $d$ quark; and $(\sqrt{2})^3$ is due to the normalization of the pion BSA.

\subsection{Elements in contact interaction}\label{sec::ci}
In this section, we briefly explains all the elements appearing in the $\gamma\to3\pi$ amplitude, namely, $S(q)$, $\Gamma_\pi(P)$, $\Gamma_\mu(Q)$ and $\Sigma^F(P_4,P_3)$. All these components are the solutions of the corresponding DSEs or Bethe-Salpeter equations (BSEs), where DSEs and BSEs are consistently truncated by the symmetry preserving MRL truncation introduced in Ref.\,\cite{Xing:2021dwe}.
 
The core idea of CI is to replace the fully dressed gluon propagator $D_{\mu\nu}(q)$, in the relevant DSEs and BSEs, with a momentum independent one\,\cite{Gutierrez-Guerrero:2010waf}:
\begin{equation}\label{eqn::CI}
    g^2D_{\mu\nu}(q)\rightarrow \frac{1}{m_G^2}\delta_{\mu\nu}.
\end{equation}
Here $m_G$ serves as gluon mass-scale. The gluon propagator would enter in the DSE for the quark propagator, as well as the BSE for the meson and QPV, among others. For the sake of brevity, we only display the inhomogeneous BSE for the QPV, which is written as:
\begin{eqnarray}\label{eqn::pion}
    \Gamma_{\mu}(P)&=&\gamma_\mu-\frac{4}{3m_G^2}\int_q \gamma_\alpha S(q)\Gamma_{\mu}(P)S(q-P)\gamma_\alpha\nn
    &+&\frac{4\xi}{3m_G^2}\int_q \tilde{\Gamma}_j S(q)\Gamma_{\mu}(P)S(q-P)\tilde{\Gamma}_j\,.
\end{eqnarray}
The first line above define the ladder truncation, while the second line in \Eqn{eqn::pion} contains the non-ladder (NL) pieces, $\tilde{\Gamma}_j=\left\{I_4,\gamma_5,\frac{i}{\sqrt{6}}\sigma_{\al\be}\right\}$; $\xi$ controls the relative strength between the ladder and NL contributions, such that $\xi=0$ recovers the traditional ladder truncation. Thus, under the approximation of Eq.\,\eqref{eqn::CI}, the relevant equations of motion would present logarithmic and quadratic divergences, so they must be regularized. A sensible regularization procedure imposes very compact forms for  the quark propagator, meson BSA and QPV. In particular, within the MRL truncation, these can be expressed as:
\begin{eqnarray}
    \!\!\!\!\!\!&&S^{-1}(p)=i \gamma \cdot p+M\,,\\
    \!\!\!\!\!\!&&\Gamma_{\pi}(P)=i\gamma_5 E_\pi(P)+ \frac{\gamma_5 \gamma \cdot P}{M} F_\pi(P)\;,\\
    \!\!\!\!\!\!&&\Gamma_{\mu}(Q)=\gamma^L_\mu f_L(Q^2)+\gamma^T_\mu f_T(Q^2)+\frac{\sigma_{\mu\nu}Q_\nu}{M} f_A(Q^2)\,,\label{eq::QPV1}
\end{eqnarray}
where $\gamma_\mu^T=\gamma_\mu-\frac{\slashed{Q}Q_\mu}{Q^2}$ and  $\gamma_\mu^L=\gamma_\mu-\gamma_\mu^T$. 
As noted, the mass function $M$ is independent of the quark momentum. The pion BSA scalar functions $E_\pi$ and $F_\pi$ depends only on the total pion momentum $P$, which are constants when pion is on-shell. Similarly, the dressing functions $f_{L,T,A}$ characterizing the QPV are also dependent only on the total photon momentum $Q$. The fact of not depending on relative moments is a characteristic of the CI model. The procedure for obtaining all these scalar functions is detailed in Ref.~\cite{Dang:2023ysl}, which follows the symmetry preserving regularization procedure developed in Ref.~\cite{Xing:2022jtt}. It is worth noting that $f_A(Q^2) \neq 0$ in Eq.\,\eqref{eq::QPV1} is obtained only if $\xi\neq0$.  That is, the component related to the quark AMM (in turn closely connected to DCSB), only manifests in the MRL truncation.

\begin{figure}[ht]
\includegraphics[width=8.6cm]{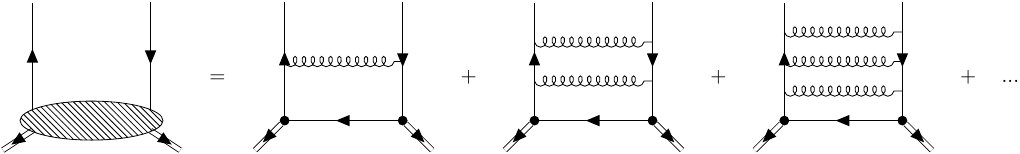}
\caption{Graphical representation of the self energy part of the $\pi-\pi$ scattering in the ladder approximation. The solid, double solid and spring lines denote quark, the pion and gluon; the solid circle and shaded ellipse denote the pion BSA and the self energy part of the $\pi-\pi$ scattering amplitude.}
\label{fig::sigma}
\end{figure}

The last element is the self-energy part of the $\pi-\pi$ scattering amplitude. In the ladder approximation, it is just the sum of infinite set of gluons exchanging diagrams, see Fig.~\ref{fig::sigma}. Hence it is equivalently described by an inhomogeneous BSE \cite{Xing:2022jtt,Cotanch:2003xv}; notably, within the MRL truncation, it is straightforward to derive it:
\begin{eqnarray}\label{eqn::sigma}
    \Sigma^F(P_1,P_2)=&-&\frac{4}{3m_G^2}\int_q \gamma_\alpha S(q_1)F(q,P_1,P_2)S(q_2)\gamma_\alpha\nn
    &+&\frac{4\xi}{3m_G^2}\int_q \tilde{\Gamma}_j S(q_1)F(q,P_1,P_2)S(q_2)\tilde{\Gamma}_j,\nn
\end{eqnarray}
where $q_1=q+P_1$, and $q_2=q-P_2$. 
%The first line above define the ladder truncation, while the second line in \Eqn{eqn::sigma} contains the non-ladder pieces, $\tilde{\Gamma}_j=\left\{I_4,\gamma_5,\frac{i}{\sqrt{6}}\sigma_{\al\be}\right\}$. 
It is worth recalling that with new structures that incorporate the MRL truncation, the relevant symmetries continue to be satisfied and the vector channels are favorably modified\,\cite{Xing:2021dwe}. Clearly, Eq.\,\eqref{eqn::sigma} exhibits a resemblance with Eq.\,\eqref{eqn::pion}, so taking $\xi=0$ would recover the result of the ladder approximation\,\cite{Xing:2022mvk}. In this case, $F(q,P_1,P_2)$ is the fully dressed $\pi-\pi$ scattering amplitude satisfying
\begin{equation}
F(q,P_1,P_2)=F_0(q,P_1,P_2)+\Sigma^F(P_1,P_2)\,,
\end{equation}
with the bare $\pi-\pi$ amplitude being
\begin{eqnarray}
F_0(q,P_1,P_2)=\Gamma_{\pi}(P_1)S(q)\Gamma_{\pi}(P_2)\,.
\end{eqnarray}
Note that the self energy $\Sigma^F(P_1,P_2)$ is independent of the quark momentum $q$, which is the feature of the CI model. The general structure of  $\Sigma^F(P_1,P_2)$ can be then decomposed as
\begin{equation}
	\Sigma^F(P_1,P_2)=\sum^4_{i=1} t_iT_i\,,
\end{equation}
where $T_i$ is a set of orthogonal basis
\begin{equation}\label{eqn::basis}
T_i=\left\{\mathbbm{1},\,\frac{-i}{M}\slashed{K},\,\frac{-i}{M}\slashed{Z},\,\frac{i}{M^2}\sigma_{\mu\nu}Z_\mu K_\nu\right\}\,,
\end{equation}
with $K=(P_1-P_2)/2$, $Z=-(P_1+P_2)$. The Kinematic relations entail $K\cdot Z=0$, $K^2=-m_\pi^2-Z^2/4$. Thus the dressing scalar functions $t_i$ can be written in terms of $Z^2$, \ie, $t_i=t_i(Z^2)$. Moreover, under the charge conjugation symmetry, $t_3(Z^2)=0$. The rest of the scalar functions can be formally written as follows by solving \Eqn{eqn::sigma}
\begin{eqnarray}
    t_1&=&\frac{b_1}{1-f_{11}},\\
    t_2&=&\frac{b_2(1-f_{44})+b_4f_{24}}{(1-f_{44})(1-f_{22})-f_{24}f_{42}},\\
    t_4&=&\frac{b_4(1-f_{22})+b_2f_{42}}{(1-f_{44})(1-f_{22})-f_{24}f_{42}},
\end{eqnarray}
where $b_i$, $f_{ik}$ ($i,k=1,2,4$) are also functions of $Z^2$, and are defined as
\begin{eqnarray}
b_{i}&=&-\frac{4}{3m_G^2}\mathcal{N}_i \text{tr}\int_q T_i S(q_1)F_0(q,P_1,P_2)S(q_2)\,,\\
f_{ik}&=&-\frac{4}{3m_G^2}\mathcal{N}_i \text{tr}\int_q T_i S(q_1)T_kS(q_2)\,,\\
\mathcal{N}_i&=&\frac{1}{\text{tr}(T_iT_i)}\frac{\gamma_\alpha T_i\gamma_\alpha -\xi\tilde{\Gamma}_j T_i\tilde{\Gamma}_j}{T_i}\,.
\end{eqnarray}
%We close this section with a few comments on the dressing functions $t_i$ in MRL truncation: 
It is important to note that the dressing functions of the scattering amplitude exhibit scalar and vector meson poles. Furthermore, the functions $t_{2/4}$ have close relations with the QPV dressing functions $f_{T/A}$, such that
\begin{eqnarray}
    t_{2}(Q^2)|_{b_2=1,b_4=0}&=&f_{T}(Q^2),\\
    t_{4}(Q^2)|_{b_2=1,b_4=0}&=&f_{A}(Q^2).
\end{eqnarray}
While the dressing $t_1$ is related to the quark-scalar vertex in an analogy, and it is in fact crucial to reproduce the low-energy prediction for the pion $D$-term \cite{Xing:2022mvk}.

\section{chiral trace and regularization}\label{sec::chiraltrace}
In this section, we briefly explain how we compute Eqs.~(\ref{eqn::box}-\ref{eqn::scat2}), which contain the anomalous chiral trace; namely, a Dirac trace that incorporates odd number of $\gamma_5$. The regularization of a chiral trace is delicate and hence should be carefully handled to avoid mathematical inconsistencies. Among many regularization schemes, the Breitenlohner-Maison/’t Hooft-Veltman (BMHV) approach \cite{tHooft:1972tcz,Breitenlohner:1975hg} in dimensional regularization is well suitable in anomaly related processes.
The $\gamma_5$ problem in dimensional regularization is intractable and well known \cite{Korner:1991sx,Jegerlehner:2000dz,Belusca-Maito:2023wah}. The following three properties cannot be satisfied at the same time in D-dimensions, \ie, cyclicity of the trace; the anti-commutation relation $\{\gamma_5,\gamma_\mu\}=0$; $\text{tr}(\gamma_5\gamma_a \gamma_b \gamma_c\gamma_d)\neq 0$.
The BMHV scheme gives up the anti-commutativity of $\gamma_5$ to ensure mathematical consistency \cite{Breitenlohner:1975hg}, while the price is that the axial vector Ward-Green-Takahashi identity (aWGTI) is violated. However, in some sense, it is the violation of this identity in the chiral trace that induces the chiral anomaly.
Unfortunately, the dimensional regularization is in general not suitable for effective theories. In the following we try to build a mapping between the dimensional regularization and the regularization developed in Ref.~\cite{Xing:2022jtt}, which is based upon the Schwinger's proper-time methon and turns out to be more suitable for the CI model, with a simple but significant example.

To start, it is helpful to introduce the so-called one-fold irreducible loop integrals (ILIs) in Ref.~\cite{Wu:2002xa}.
\begin{eqnarray}
I_{-2\alpha}(\mathcal{M}^2)&=&\int_{q}\frac{1}{(q^2+\mathcal{M}^2)^{\alpha+2}}\,,\nn
I^{\mu\nu}_{-2\alpha}(\mathcal{M}^2)&=&\int_{q}\frac{q_{\mu}q_{\nu}}{(q^2+\mathcal{M}^2)^{\alpha+3}}\,,\nn
I^{\mu\nu\rho\sigma}_{-2\alpha}(\mathcal{M}^2)&=&\int_{q}\frac{q_{\mu}q_{\nu}q_{\rho}q_{\sigma}}{(q^2+\mathcal{M}^2)^{\alpha+4}},
\end{eqnarray}
where $\mathcal{M}$ is a function of Feynman parameters, external momenta and the corresponding mass scales, with $\alpha=-1,0,1,\cdots$. Here $\alpha=-1$ represents quadratically divergent integrals, and $\alpha=0$ represents logarithmically divergent integrals.
In Ref.~\cite{Xing:2022jtt}, the regularized ILIs, which is based on Schwinger's proper time method, satisfy the following consistency conditions
\begin{eqnarray}
I^{\mu\nu R}_{-2\alpha}(\mathcal{M}^2)&=&\frac{\Gamma(\alpha+2)}{2\Gamma(\alpha+3)}\delta_{\mu\nu}I^R_{-2\alpha}(\mathcal{M}^2)\,,\\
I^{\mu\nu\rho\sigma R}_{-2\alpha}(\mathcal{M}^2)&=&\frac{\Gamma(\alpha+2)}{4\Gamma(\alpha+4)}S_{\mu\nu\rho\sigma}I^R_{-2\alpha}(\mathcal{M}^2)\,,
\end{eqnarray}
where $S_{\mu\nu\rho\sigma}=\delta_{\mu\nu}\delta_{\rho\sigma}+\delta_{\mu\rho}\delta_{\sigma\nu}+\delta_{\mu\sigma}\delta_{\nu\rho}$ is the totally symmetric tensor. The label $R$ denotes the regularization scheme in Ref.~\cite{Xing:2022jtt}, and the integral $I^R_{-2\alpha}(\mathcal{M}^2)$ is defined as follows:
\begin{equation}
    I^R_{-2\alpha}(\mathcal{M}^2) = \int_{1/\Lambda_{uv}^2}^{1/\Lambda_{ir}^2} d\tau \frac{\tau^{\alpha-1}}{\Gamma(\alpha+2)}\frac{e^{-\tau \mathcal{M}^2}}{16\pi^2}\,,
\end{equation}
where $\Lambda_{ir,uv}$ are the infrared and ultraviolet regulators, respectively. While $\Lambda_{uv}$ plays a dynamical role and cannot be removed from the theory, the infrared regulator $\Lambda_{ir}$ is introduced in the CI scheme to prevent quark production thresholds and thus producing a picture compatible with confinement\,\cite{Gutierrez-Guerrero:2010waf}.

In a normal case (the counterpart to this is the chiral trace), the scheme $R$ works fine since gauge and chiral symmetries are preserved due to the consistency conditions. Actually, dimensional regularization also satisfies the consistency conditions so that symmetry is also preserved therein. However, things become anomalous when the chiral trace is regularized. The subtle regularization of this case is illustrated below. Let's consider the following integral
\begin{equation}\label{eqn::ctintegral}
    A=\int_q \frac{CT}{(q^2+\mathcal{M}^2)^{3}},
\end{equation}
where the numerator is a chiral trace
\begin{equation}
    CT:=-\frac{1}{4}\text{tr}(\gamma_5\slashed{q}\gamma_a\gamma_b\gamma_c\gamma_d\slashed{q}).
\end{equation}
There are many ways to evaluate this chiral trace. Here we focus on two options. The first option is to use the dirac algebra and commute $\slashed{q}$ with $\gamma_{a,b,c,d}$ until the two $\slashed{q}$ meet. This option leads to
\begin{eqnarray}\label{eqn::ct1}
    CT_{1}&=&2q_\mu(q_a \epsilon_{bcd\mu}-q_b \epsilon_{acd\mu}+q_c \epsilon_{abd\mu}-q_d \epsilon_{abc\mu})\nn
    &+&q^2\epsilon_{abcd}.
\end{eqnarray}
Then under the scheme $R$, the regularized \Eqn{eqn::ctintegral} is
\begin{eqnarray}
    A^{R}_{1}&=&\left(-8\frac{I_{0}^R(\mathcal{M}^2)}{4}+I_{0}^R(\mathcal{M}^2)-\mathcal{M}^2 I_{-2}^R(\mathcal{M}^2)\right)\epsilon_{abcd},\nn
             &=&-\left(I_{0}^R(\mathcal{M}^2)+\mathcal{M}^2 I_{-2}^R(\mathcal{M}^2)\right)\epsilon_{abcd}.
\end{eqnarray}
The second option is to use cyclicity of trace and commute $\slashed{q}$ with $\gamma_5$. Here comes the delicate part, inspired by the BMHV scheme, we assume that the anti-commutativity of $\gamma_5$ with the loop momentum $\slashed{q}$ do not hold in the case of chiral trace, even though the space-time dimension is $4$ in the scheme $R$. One then obtains
\begin{equation}\label{eqn::ct2}
    CT_{2}=-q^2\epsilon_{abcd}+\delta,
\end{equation}
where $\delta$ is a result of $\{\gamma_5,\slashed{q}\}\neq0$, and $\delta$ vanishes at the level of integrand but gives a non-zero contribution $\Delta$ after integration (regularization). Then the regularized \Eqn{eqn::ctintegral} can be then written as
\begin{eqnarray}
    A^{R}_{2}&=&-\left(I_{0}^R(\mathcal{M}^2)-\mathcal{M}^2 I_{-2}^R(\mathcal{M}^2)\right)\epsilon_{abcd}+\Delta.
\end{eqnarray}
Mathematical consistency requires that $CT_{1}=CT_{2}$ and $A^{R}_{1}=A^{R}_{2}$. Comparing these two options, we can prove that $\delta=0$, at integrand level, with the famous Schouten identity
\begin{equation}
    g_{\mu \nu}\epsilon_{abcd}=g_{\mu a}\epsilon_{\nu bcd}+g_{\mu b}\epsilon_{a\nu cd}+g_{\mu c}\epsilon_{ab\nu d}+g_{\mu d}\epsilon_{abc\nu}.
\end{equation}
While for the regularized \Eqn{eqn::ctintegral}, we find that
\begin{equation}
    \Delta=-2\mathcal{M}^2 I_{-2}^R(\mathcal{M}^2)\epsilon_{abcd}\,.
\end{equation}
This $\Delta$, originated from $\{\gamma_5,\slashed{q}\}\neq0$ in the chiral trace, induces the well-known chiral anomaly. On the other hand, $\{\gamma_5,\slashed{q}\}\neq0$ also means the violation of the aWGTI. Hence, chiral anomaly and the violation of aWGTI would be interconnected via the chiral trace.

In 4-dimensional space-time, it is hard to define the explicit form of $\delta$, but its meaning becomes clear in dimensional regularization (here we only extend the loop momentum $q$ to $D$ dimensions). In dimensional regularization with BMHV scheme, the chiral traces $CT_{1,2}$ have the same form as in \Eqn{eqn::ct1} and \Eqn{eqn::ct2} except that the 4-dimensional loop momentum $q$ is extend to $D$ dimensions $q_D$ and, more importantly, 
\begin{eqnarray}
    \delta=2q_{D-4}^2\epsilon_{abcd},
\end{eqnarray}
where $q_{D-4}$ is the $D-4$ part of the loop momentum $q_D$ such that $q_D=q+q_{D-4}$. In dimensional regularization, we have
\begin{eqnarray}
    &&\int_{q}^{DR}\frac{q_D^2}{(q_D^2+\mathcal{M}^2)^{\alpha+3}}=\frac{D}{2(\alpha+2)}I^{DR}_{-2\alpha}(\mathcal{M}^2),\\
    &&\int_{q}^{DR}\frac{q^2}{(q_D^2+\mathcal{M}^2)^{\alpha+3}}=\frac{4}{2(\alpha+2)}I^{DR}_{-2\alpha}(\mathcal{M}^2),\\
    &&\int_{q}^{DR}\frac{q_{D-4}^2}{(q_D^2+\mathcal{M}^2)^{\alpha+3}}=\frac{D-4}{2(\alpha+2)}I^{DR}_{-2\alpha}(\mathcal{M}^2),
\end{eqnarray}
where $\int_{q}^{DR}\doteq\int\frac{d^Dq_D}{(2\pi)^D}$ and $I^{DR}_{-2\alpha}$ is the regularized ILIs in dimensional regularization. Subsequently, by taking $\alpha=0$ and absorbing explicit dimension parameter $D$ into $I^{DR}_{-2\alpha}(\mathcal{M}^2)$, we find that
\begin{eqnarray}
    &&\int_{q}^{DR}\frac{q_D^2}{(q_D^2+\mathcal{M}^2)^{3}}=I^{DR}_{0}(\mathcal{M}^2)-\mathcal{M}^2I^{DR}_{-2}(\mathcal{M}^2),\\
    &&\int_{q}^{DR}\frac{q^2}{(q_D^2+\mathcal{M}^2)^{3}}=I^{DR}_{0}(\mathcal{M}^2),\\
    &&\int_{q}^{DR}\frac{q_{D-4}^2}{(q_D^2+\mathcal{M}^2)^{3}}=-\mathcal{M}^2I^{DR}_{-2}(\mathcal{M}^2).
\end{eqnarray}
One then obtains in dimensional regularization
\begin{equation}
    \Delta=-2\mathcal{M}^2 I_{-2}^{DR}(\mathcal{M}^2)\epsilon_{abcd}\,,
\end{equation}
and this form of $\Delta$ embodies the mathematical consistency of the BMHV scheme which implies 
\begin{equation}
    A^{DR}_{1}=A^{DR}_{2}=-\left(I_{0}^{DR}(\mathcal{M}^2)+\mathcal{M}^2 I_{-2}^{DR}(\mathcal{M}^2)\right)\epsilon_{abcd}.
\end{equation}
The last point we want to stress is that in dimensional regularization, $\delta=0$ when $D\to4$ at the level of integrand. But after integration it gives a non-zero value, as argued early in the birth of dimensional regularization \cite{tHooft:1972tcz}.

From the above example, we have illustrated why the regularization of the chiral trace is subtle and, in some sense, the origin of the chiral anomaly. In the meantime, we see that a clear mapping between dimensional regularization and regularization $R$ is established:
\begin{equation}\label{eqn::map}
    I_{-2\alpha}^{DR}(\mathcal{M}^2)\leftrightarrow I_{-2\alpha}^{R}(\mathcal{M}^2).
\end{equation}
Therefore, in computing Eqs.~(\ref{eqn::box}-\ref{eqn::scat2}), we first extend the loop momentum $q$ to $D$ dimensions, and then apply BMHV scheme. After absorbing explicit dimensional parameter $D$ into the regularized ILIs, Eqs.~(\ref{eqn::box}-\ref{eqn::scat2}) can be expressed in terms of $I_{-2\alpha}^{DR}(\mathcal{M}^2)$. The final step is to apply the mapping \Eqn{eqn::map} by replacing $I_{-2\alpha}^{DR}(\mathcal{M}^2)$ with $I_{-2\alpha}^{R}(\mathcal{M}^2)$.

\section{results}\label{sec::results}
%When evaluating the chiral trace (meaning a dirac trace that incorporates odd number of $\gamma_5$), one faces trace ambiguities. In this paper, we evaluate the chiral trace with the help of dimensional regularization. In dimensional regularization, the BMHV scheme is a mathematically consistent scheme to evaluate the chiral trace, thus we adopt this scheme and connect it with the regularizaiton described in \cite{Xing:2022jtt}. Some discussions about the chiral trace and the connection between dimensional regularization and the regularizaiton in \cite{Xing:2022jtt} are presented in Appendix.~\ref{app::chiraltrace}.
\subsection{Comparison with low energy therorem}\label{sec::lowenergy}
In this section, we compare our results with the low energy theorem. To address this observation, let us consider the $\xi=0$ case in the chiral limit, $m_{\pi}=0$, and soft-pion limit $s=t=u=0$. After the careful treatment of the chiral trace and regularizaiton, it turns out that the divergences cancel out and all the amplitudes in Eqs.~(\ref{eqn::box}-\ref{eqn::scat2}) contain only convergent integrals. In principle, the cutoff should not be removed in an effective theory, meaning that convergent integrals should also be regularized. Nonetheless, for the sake of comparison with the low energy theorem, we decide first not to regularize those integrals and to completely remove the cutoffs.\par
After above clarifications, the computed amplitude for the box diagram might be expressed as
\begin{equation}
\label{eq:Abox1}
    A^{box}(0,0,0)=e\frac{E_{\pi } \left(E_{\pi }^2-6 E_{\pi } F_{\pi }+6 F_{\pi }^2\right)}{4 \pi ^2 M^3}\,.
\end{equation}
We now proceed to examine the pseudovector component of the pion BSA $F_\pi$. By setting $F_\pi=0$, the box diagram would reproduce the well known low energy prediction $A_0^{3\pi}$ by using the Goldberger-Treiman relations in the chiral limit \cite{Dang:2023ysl}, entailing
\begin{equation}\label{eqn::gt}
    f_\pi E_\pi=M\,.
\end{equation}
However, in the framework of a self-consistent CI model (or in general in a symmetry-preserving treatment of the corresponding DSEs and BSEs), the pseudovector structure is naturally generated and cannot be zero\,\cite{Maris:1998hc}. Therefore, we define the relative ratio of pion BSA $\mathcal{R}_\pi\doteq F_\pi/E_\pi$, and then the contribution from box diagram is re-expressed as
\begin{equation}
    A^{box}(0,0,0)=(1-6 \mathcal{R}_{\pi }+6 \mathcal{R}_{\pi }^2) A_0^{3\pi}\,,
\end{equation}
where the chiral limit value of $\mathcal{R}_\pi$ is
\begin{equation}\label{eqn::rpi}
    \mathcal{R}_\pi=\frac{4f_\pi^2}{3m_G^2}\frac{1}{2N_c}\,.
\end{equation}
The derivation of \Eqn{eqn::rpi} is presented in Appendix.~\ref{app::bsaratio}.

Let us now examine the $\pi-\pi$ scattering diagram. According to the structures in \Eqn{eqn::basis}, it is seen that the scattering diagram corresponds to the $\pi\to2\gamma$ triangle diagram, hence producing
\begin{equation}
    A^{scat}(0,0,0)=-6\frac{t_2(0)}{M}\frac{Q_u+Q_d}{Q_u^2+Q_d^2} A^{\pi}_0\,,
\end{equation}
where according to \cite{Xing:2022mvk} and the conservation of electric charge
\begin{equation}
    t_2(0)=-\frac{4M}{3m_G^2}\frac{1}{2N_c}\,.
\end{equation}
In terms of $\mathcal{R}_\pi$, the contribution from scattering diagram can be written as
\begin{equation}
    A^{scat}(0,0,0)=6 \mathcal{R}_\pi A^{3\pi}_0\,.
\end{equation}
Thus, the total form factor has the form:
\begin{eqnarray}\label{eqn::result}
    A^{3\pi}(0,0,0)&=&A^{box}(0,0,0)+A^{scat}(0,0,0),\nn
    &=&(1+6 \mathcal{R}_\pi^2)A_0^{3\pi}.
\end{eqnarray}
We can notice that there is a correction term to the low energy prediction for the $\gamma\to3\pi$ anomaly. This is not the case for the $\pi\to2\gamma$ process, where $F_\pi$ does not contribute to the triangle diagram for the chiral limit pion and on-shell photons.  Hence, one obtains exactly the low energy prediction $A^{\pi}_0$ if convergent integrals are not regularized\,\cite{Gutierrez-Guerrero:2010waf,Dang:2023ysl}.\par
Finally, in the CI model interpretation, the correction term $6 R_\pi^2$ to the $\gamma\to3\pi$ anomaly is relevant due to its connection to the relative ratio of the pion BSAs. To further understand this correction, we find that after taking into account axial vector field in the WZW action, $ \mathcal{R}_\pi$ could be understood as the shifting coefficients of the axial vector field
\begin{equation}
    a'_\mu=a_\mu-\mathcal{R}_\pi \partial_\mu \pi\,,
\end{equation}
here $a_\mu$ and $\pi$ are the axial vector and pseudoscalar field, respectively, and $a'_\mu$ is the shifted field. The details can be found in Ref.~\cite{Li:2004ku,Osipov:2020hjd}.

\subsection{Numerical results}\label{sec::momentum}
Having provided results related to the chiral anomaly and, in particular, accounting for the origin and magnitude of the violation of the low energy theorem, we now proceed to discuss the  influence of the different components entering the calculation: the regularization procedure, the NL pieces of the MRL truncation, and the components of the pion BSA. For consistency, we use the same parameters as in Ref.~\cite{Dang:2023ysl}. The computed masses, decay constants and the normalized BS amplitudes of the pion meson, as well as the mass function of dressed quark, are reported in Table ~\ref{tab:masscball}. 

\begin{table}[htpb]
%\vspace*{-3mm}
\caption{\label{tab:masscball} Computed pion static properties in the case of the chiral limit and the physical pion mass. The model parameters: $m_G=0.132\,\GeV$, $\Lambda_{uv}=0.905\, \GeV$ and  $\Lambda_{ir}=0.24\,\GeV$. Mass units in $\GeV$.}
\setlength{\tabcolsep}{1.7mm}{
\begin{tabular}{c|ccccc}
\hline
% $m$&$m_G$&$\Lambda_{uv}$&$\Lambda_{ir}$&$M$ &$m_{\pi}$  & $f_{\pi}$  &$E_{\pi}$ & $F_{\pi}$\\
  &$M$ &$m_{\pi}$  & $f_{\pi}$  &$E_{\pi}$ & $F_{\pi}$\\
\hline
chiral &0.358 & 0 & 0.100 &3.566 &  0.458 \\
\hline
physical&0.368 & 0.140 & 0.101 &3.595 &  0.475 \\
\hline
\end{tabular}}
%\vspace*{-3mm}
\end{table}
We highlight the value of $\xi=0.151$, a choice that indicates the NL components of the MRL truncation are activated. In the case of the $\pi\to2\gamma$ decay, it has been seen that these structures generated in the MRL truncation mimic the complex dynamics beyond the cutoff, therefore setting $\xi=0.151$ to faithfully reproduce the related anomaly which otherwise would be underestimated~\cite{Dang:2023ysl}. Therefore, we will see 
whether this contribution, with the same value of $\xi=0.151$, would be sufficient to simultaneously reproduce the anomalies related to the $\pi\to2\gamma$ and $\gamma\to3\pi$ processes. Namely, one should check whether the regularized numerical results with $\xi=0.151$ can match \Eqn{eqn::result} in the chiral limit. 

For that purpose, let us consider the chiral limit value of the $\gamma\pi\pi\pi$ amplitude, $A^{3\pi}_0$, which provides a useful normalization. We therefore define the function
\begin{equation}\label{eqn::amplitude}
    \tilde{A}^{3\pi}(0,0,0)\doteq A^{3\pi}(0,0,0)/A^{3\pi}_0.
\end{equation}
The numerical results for $\tilde{A}^{3\pi}(0,0,0)$ in various situations are reported in Table ~\ref{tab:check}: $\xi=0$ in the regularized and not regularized cases, as well as $\xi=0.151$ in the regularized case. A few points can be drawn from the results: Firstly, comparing the first two rows, we can notice again that the value of the anomaly is reduced due to the cutoff effect, which would imply that something is being left out due to the presence of the cutoff. Secondly, the second and third row shows that the "box+scat" results are nearly the same. This means the additional structure generated by the MRL truncation can indeed compensate the cutoff effect for the $\gamma\to3\pi$ process with the same parameter $\xi=0.151$ determined from $\pi\to2\gamma$ process. This result validate the idea proposed in Ref.~\cite{Dang:2023ysl}. On the other hand, unlike "box+scat", there is a mismatch in the second and the third row for "box (only E)" and "box". This can be explained from the fact that $F_\pi$ and/or the scattering amplitude are not included so that the framework is not complete. As a consequence, the additional structures generated by the MRL truncation fail to compensate the cutoff effect.

\begin{figure}[t]
\includegraphics[width=8.6cm]{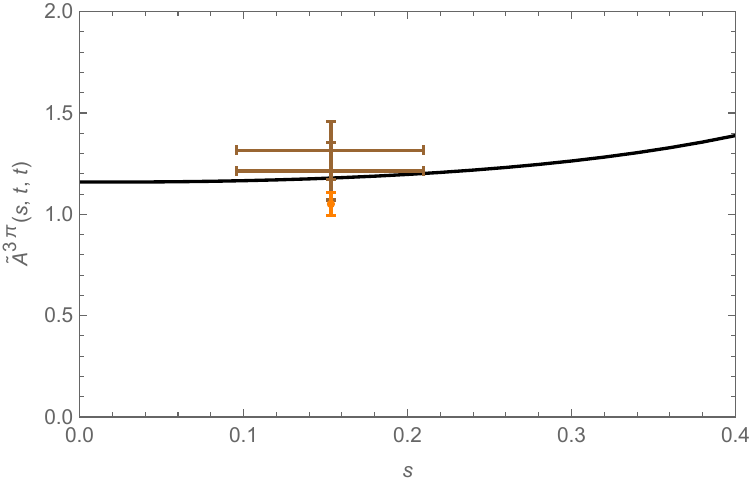}
\caption{Comparison of $\tilde{A}^{3\pi}(s,t,t)$ with available data. Solid curve -  momentum dependence of the form factor$\tilde{A}^{3\pi}(s,t,t)$. Experimental data points are from Refs. \cite{Antipov:1986tp,Ecker:2023qae}, Brown polygons and Orange disks, respectively.
%Experimental data point is from Ref. \cite{Antipov:1986tp},Brown polygons, and Lattice data point is from Ref. \cite{Ecker:2023qae}, Orange disks 
}
\label{fig::tffstt}
\end{figure}

\begin{table}[ht]
%\vspace*{-3mm}
\caption{\label{tab:check} Results for $\tilde{A}^{3\pi}(0,0,0)$. Here "only E" means that we use $\Gamma_{\pi}(P)=i\gamma_5 E_\pi(P)$ in the calculation of all the pion related quantities, in which case $f_\pi=0.116\GeV$; "box" means $A^{3\pi}=A^{box}$, and "box+scat" refers to $A^{3\pi}=A^{box}+A^{scat}$. The $A^{box}$ and $A^{scat}$ contributions are defined through Eqs.~(\ref{eqn::box}-\ref{eqn::scat2}).  All calculations are performed in the chiral limit.}
\setlength{\tabcolsep}{1.7mm}{
\begin{tabular}{l|cccc}
\hline
&box (only E) &box  &box+scat\\
\hline
$\xi=0$, regularized  &0.640 & 0.193 &0.769 \\
\hline
$\xi=0$, not regularized  &1 & 0.328 &1.099 \\
\hline
$\xi=0.151$, regularized&0.725 & 0.231 &1.105 \\
\hline
\end{tabular}}
%\vspace*{-3mm}
\end{table} 
We also calculate the momentum dependent form factor $A^{3\pi}(s,t,u)$ at physical pion mass. Direct computation indicates that 
\begin{eqnarray}
   f^{box}\propto f_{T,A},\ f^{scat}\propto f_{T,A}\times t_{2,4}.
\end{eqnarray}
As mentioned in Sec.~\ref{sec::ci}, the dressing functions $f_{T,A}$ and $t_{2,4}$ posses vector meson poles. So that the pole structure of the form factor can be characterized as
\begin{eqnarray}
    A^{box}(s,t,u)\!\!&\sim&\!\!\frac{1}{Q^2+m_\rho^2},\\
    A^{scat}(s,t,u)\!\!&\sim&\!\!\frac{1}{Q^2+m_\rho^2}\left(\frac{1}{m_\rho^2-s}+\frac{1}{m_\rho^2-t}+\frac{1}{m_\rho^2-u}\right),\nn
\end{eqnarray}
which produces a picture congruent with the phenomenological vector meson dominance (VMD) parametrization. Our numerical results are shown in Fig.~\ref{fig::tffstt}. The solid line is our calculated result for
 the $\gamma\pi\pi\pi$ amplitude $\tilde{A}^{3\pi}(s,t,t)$, \Eqn{eqn::amplitude}, as a function of the Mandelstam variable $s$. The kinematics is such that $Q^2=0$ and $u=t$, and all pions are on shell \ie \,$P_{2}^2=P_{3}^2=P_{4}^2=-m_\pi^2$. We have demonstrated that in the chiral limit $A^{3\pi}_0$ is in fact independent of the model parameters. The evolution of $A^{3\pi}(s,t,u)$ with $s$ does depend on the model parameters, even in the chiral limit. We see that in the small $s$ region, roughly below $0.1\,\GeV^2$, the amplitude merely changes with respect to $s$, while the vector meson pole leads to sensible variations of the amplitudes when $s>0.1\,\GeV^2$.
 
 Our results are comparable with that obtained in other models \cite{Alkofer:1995jx,Holstein:1995qj,Benic:2011rk,Hoferichter:2012pm,zhevlakov2017transition,bistrovic2000quark}. The comparison with existing experimental data points \cite{Antipov:1986tp,Ecker:2023qae} is also displayed in Fig.~\ref{fig::tffstt}. Since the data reported in  Ref.\,\cite{Antipov:1986tp} is significantly higher than the low energy prediction, it has raised some concerns.  However, considering the experimental errors and the predictions of our model, this data point doesn’t seem implausible. Whereas the recent data point \cite{Ecker:2023qae} has a reduced error band and a fair agreement with our prediction. 
 
 Finally, we close this section with by discussing the impact of the quark mass effect on the $\gamma\to3\pi$ anomaly. Although the correction in \Eqn{eqn::result} is obtained in the chiral and soft pion limits, the discrepancy from low energy theorem and experiment can also be explained as a beyond chiral limit effect. It has been reported in Refs.~\cite{Hoferichter:2012pm,Hoferichter:2017ftn} that a quark mass renormalization effect indeed increases the chiral anomaly by about $7\%$ through a resonance-saturation estimate~\cite{Bijnens:1989ff}. Herein we also choose some kinematic configurations to explore the quark mass effect. The first configuration is to simply set $s=t=u=0$, which leads to $Q^2=3m_\pi^2$;  the second configuration is to set $Q^2=0$ and $s=u=t=m_\pi^2$; a the third configration corresponds to setting $Q^2=s=0$ and $u=t=\frac{3m_\pi^2}{2}$. In the chiral limit, with vanishing current quark mass, $m_\pi=0$ and all configurations recover\,\Eqn{eqn::result}. A finite current quark mass leads to a physical pion mass $m_\pi=0.140$ GeV, as presented in Table.~\ref{tab:masscball}. Within our model and input parameters, we find that for the first configuration, the amplitude $A^{3\pi}(0,0,0)$ reduces by about $6\%$ because of the current quark mass effect, which qualitatively agrees with the early DSE calculations~\cite{Cotanch:2003xv}. As for the rest two configurations, where the photon is on-shell, both the amplitudes $A^{3\pi}(s=m_\pi^2,s,s)$ and $A^{3\pi}(s=0,t,t)$ increase by around $2\%$ due to the finite current quark mass, which is relatively small compared to the $\sim10\%$ correction in \Eqn{eqn::result}. Nonetheless, in comparison with the rather accurate experiment result for the $\pi\to2\gamma$ transition, the data accuracy for the $\gamma\to3\pi$ process is still not satisfying, leaving room for different theoretical interpretations.

\section{SUMMARY}\label{sec::summary}
In this work we have calculated the $\gamma\to3\pi$ form factor in DSEs formalism, in specific, the CI model embedded in the so-called MRL truncation. The amplitudes are carefully computed in the way described in Sec.~\ref{sec::chiraltrace}. We find that if we only consider the leading structure of of the pion \ie, $i\gamma_5$, the low energy theorem is reproduced. However, this is just a spurious result because the framework is incomplete. A full, self-consistent calculation for this form factor requires that all structures of the pion BSA are included and the $\pi-\pi$ scattering effect must be considered in addition to the box diagram. Subsequently, our primary result, \Eqn{eqn::result}, indicates that there should be a correction to the low energy prediction, making the anomaly about $10\%$ larger than $A^{3\pi}_0$ with our input parameters.\par It is also important to note that the NL pieces contained within MRL truncation, generate additional structures in the QPV and the $\pi-\pi$ scattering amplitude. In particular, the QPV features a quark AMM contribution whose origin can be traced back to DCSB, and the $\pi-\pi$ scattering amplitude develops an analogous $T_4$ component that can make up the lost contribution to the $\gamma\to3\pi$ anomaly by using the parameters determined by the $\pi\to2\gamma$ anomaly. The ideas proposed in\,\cite{Dang:2023ysl} are in some degree validated by this simultaneously reproducing the chiral anomaly of the $\pi\to2\gamma$ and $\gamma\to3\pi$. This encouraging result may be used in investigating the momentum dependence of anomaly related process for effective theories. Finally, we give the momentum dependent form factor which is comparable with existing experimental data. The theoretical study of these processes is also relevant to issues related to the anomalous magnetic moment of the muon, where the precise determination of the corresponding couplings, and its running with the photon momenta, would be crucial\,\cite{Knecht:2001qf, Hoferichter:2019mqg,Hoferichter:2018kwz,Hoferichter:2018dmo}. \par% and other theoretical results. \par
%It is also important to note that the additional structure created by the truncation of the MRL, i.e., the amplitude modulus (AMM) in QPV and the scattering amplitude (T4) of the QPV, can account for the lost contribution to pipi anomaly using the parameter defined by the pi anomaly.
%It is also important to note that the additional structure created by the MRL truncation—that is, the AMM in the QPV and T4 in the scattering amplitude—can use the parameter found by the pi anomaly to make up for the lost contribution to the pipi anomaly.
So far, the distinctions between theories have not been made very clear. There are some unanswered questions. Especially, is there a deviation from the low energy prediction $A^{3\pi}_0$? If the deviation is confirmed, how can it be understood and explained? The answer to these questions is of great significance in understanding QCD.
Therefore, more precise experimental data is urged to make the situation clear.

\hspace*{\fill}\ 
\begin{acknowledgments}
Work supported by National Natural Science Foundation of China (grant no. 12135007). K.~R.~is supported by the Spanish MICINN grant PID2019-107844-GB-C2, and regional Andalusian project P18-FR-5057.
\end{acknowledgments}
%\newpage
\appendix
\section{derivation of \texorpdfstring{$\mathcal{R}_\pi$}{} in the chiral limit}\label{app::bsaratio}
To derive \Eqn{eqn::rpi} we start by solving the pion BSE in the chiral limit. Following a standard procedure (see Refs.~\cite{Gutierrez-Guerrero:2010waf,Dang:2023ysl}) we obtain
\begin{equation}
    \mathcal{R}_\pi=\frac{M^2 I_0^R(M^2)}{2(I_2^R(M^2)+M^2 I_0^R(M^2))}\,.
\end{equation}
In the chiral limit, $I^R_2(M^2)$ is related to the quark DSE
\begin{equation}
    M=\frac{16 M}{3m_G^2}I_2^R(M^2)\,;
\end{equation}
while $I_0^R(M^2)$ can be related with the canonical normalization condition of the pion BSA
\begin{eqnarray}
    1&=&4N_c E_\pi(E_\pi-2F_\pi) I_0^R(M^2),\nn
    &=&4N_c E_\pi^2(1-2\mathcal{R}_\pi) I_0^R(M^2)\,.
\end{eqnarray}
In conjugation with \Eqn{eqn::gt}, one finally obtains \Eqn{eqn::rpi}.

\bibliography{apstemplateNotes}
\end{document}